\documentclass[aps,prl,showpacs, twocolumn]{revtex4}
\usepackage{amsmath}
\usepackage{epsfig}
\usepackage{amssymb}
\usepackage{color}
\usepackage{graphicx}

\newcommand{\eqb}{\begin{eqnarray}}
\newcommand{\eqe}{\end{eqnarray}}

\newcommand{\bK}{{\bf K}}

\newcommand{\bq}{{\bf q}}

\newcommand{\bH}{{\bf H}}
\newcommand{\bG}{{\bf G}}
\newcommand{\bw}{{\bf w}}
\newcommand{\tbw}{\tilde{\bf w}}
\newcommand{\tb}{\tilde{b}}

\newcommand{\cP}{{\cal P}}

\def\bes{\begin{subequations}}
\def\ees{\end{subequations}}

\newcommand{\PT}{{\cal PT}}
\newcommand{\T}{{\cal T}}
\newcommand{\p}{{\cal P}}

\newcommand{\IM}{\textrm{Im}}
\newcommand{\diag}{\mbox{diag}}

\begin{document}

\title{Nonlinear modes in finite-dimensional  ${\cal PT}$-symmetric systems}

\author{D. A. Zezyulin and V. V. Konotop}
\affiliation{
Centro de F\'isica Te\'orica e Computacional and Departamento de F\'isica,   Faculdade de Ci\^encias, Universidade de Lisboa, Avenida Professor Gama Pinto 2, Lisboa 1649-003, Portugal}

\begin{abstract}

By rearrangements of waveguide arrays with gain and losses one can
simulate transformations  among  parity-time (${\cal PT}$-)
symmetric systems not affecting their pure real linear spectra.
Subject to such transformations, however, the nonlinear properties
of the systems undergo significant changes. On an example of an
%
array of four waveguides described by the discrete nonlinear
Schr\"odinger equation with dissipation and gain, we show that the
equivalence of the underlying linear spectra implies similarity of
neither structure nor stability  of the nonlinear modes in the
arrays. Even the existence of one-parametric families of nonlinear
modes is not guaranteed by the ${\cal PT}$ symmetry of a newly
obtained system. Neither the stability is directly related to the
${\cal PT}$ symmetry: stable nonlinear modes exist even when the
spectrum of the linear array is not purely real. We use graph
representation of ${\cal PT}$-symmetric networks allowing for
simple illustration of linearly equivalent networks and indicating
on their possible experimental design.
\end{abstract}


\maketitle

The effect of dissipation or gain on  dynamics of a physical
system is a fundamental issue either in classical or in quantum
theories. Optics is one of the areas where the respective models
appear naturally and are explored already for many years in
contexts of different kinds of dissipative
solitons~\cite{diss_solit}. One of a number of widely used,
fundamental and simple  models is an array of waveguides in the
presence of gain and losses~\cite{waveguide}. This model is
described 
by the discrete nonlinear Schr\"odinger  equation (DNLSE), which
is fairly general. Its applications   range from the so-called
discrete optics~\cite{disc_opt} to biophysics~\cite{biophysics}
and the meanfield theory of Bose-Einstein condensate~\cite{BEC}
(for a broad range of applications of DNLSE see
also~\cite{general}).

Recently, great interest in systems with dissipation and gain was
triggered by the discovery of the so-called parity-time ($\PT$)
potentials, which in a definite range of parameters obey purely
real spectrum~\cite{Bender}. Numerous linear physical systems for
which $\PT$ symmetry is of great relevance have been proposed.
Among them we mention non-Hermitian extension of quantum
mechanics~\cite{quantum}, electromagnetic wave propagation in  a
planar waveguide filled with active media~\cite{Muga}, and beam
propagation in optical lattices~\cite{Markis}. The phenomenon of
$\PT$ symmetry breaking has been experimentally implemented  in
optics~\cite{experiment}, where the  equations governing the
system  were earlier known as describing a unidirectional coupler,
i.e. as a particular form of the DNLSE~\cite{waveguide}.


Nonlinear $\PT$-symmetric problems  were first posed in the
context of the quantum field theory accounting for cubic
interactions~\cite{Bender_nonlin} and in guided wave
theory~\cite{Christodoulides1}. Being natural for optical
applications, the nonlinear problems received particular attention
in the context  of existence of  gap
solitons~\cite{Christodoulides1} and  defect modes~\cite{defect}
in $\PT$-symmetric lattices, as well as  in context of the
nonlinear $\PT$-symmetric couplers in
stationary~\cite{coupler_stat,Kevr2011} and
solitonic~\cite{coupler_solit} regimes. More generally,  the
nonlinearity  enriches possible statements of the problem allowing
for including the effects on nonlinear~\cite{PT-nonlin}, as well
as both linear and nonlinear~\cite{PT-lin-nonlin} $\PT$-symmetric
potentials.
%

It is known that by applying a  similarity transformation to a
given linear $\PT$-symmetric system, a new system with real
spectrum can be constructed. Thus in~\cite{Darboux} new potentials
(not necessarily $\PT$-symmetric) with real spectra were
constructed using the Darboux transformation, while
in~\cite{Mostaf2003} pseudo-Hermitian operators were introduced
and unitary equivalence  of  $\PT$-symmetric and Hermitian
operators was established. It turns out, however, that possible
mutual reductions  of Hermitian, $\PT$-symmetric,  and
pseudo-Hermitian linear operators leaving the spectrum pure real,
may introduce dramatic  changes in the properties of the
respective nonlinear systems. The analysis of such changes is the
main goal of the present Letter.

More specifically, we show that $\PT$-symmetric systems obeying
the same linear spectrum  may either have  one-parametric families
of nonlinear modes or have not. If the families exist,  stability
of the modes is essentially different for different systems, still
having the same linear spectrum.  Moreover,  stable nonlinear
modes may exist beyond the $\PT$ symmetry breaking. For a discrete
system consisting of four waveguides we  find that breaking of
$\PT$ symmetry can occur in two different ways: the linear
spectrum acquires either two complex and two real eigenvalues, or
all four  eigenvalues become complex. Finally,   we represent each
underlying  linear system by a graph, allowing one to catalog
different linearly equivalent $\PT$-symmetric systems.

We consider an array of $N$ waveguides (sites) and denote the
field in the $n$th waveguide by $q_n(z)$, where $z$ is  the
propagation distance.  If each waveguide have dissipation or gain
described by $\gamma_n$,  positive or negative, respectively, then
the field propagation  is governed by the DNLSE
 \begin{eqnarray}
\label{dynam_0}
i\dot q_n= - \sum_{m=1}^{N} K_{nm} q_m -
|q_n|^2q_n - i\gamma_nq_n,
\end{eqnarray}
where $\dot q_n=dq(z)/dz$.
Here we admit the existence
of non-local coupling among the waveguides, which is described by the coefficients $K_{nm}=K_{mn}= {K}^*_{nm}$ which  will be treated as entries of the real symmetric matrix $\bK$:  $\bK=\bK^\dag$, where $\bK^\dag$ is the Hermitian conjugate  matrix. 
It is convenient  to introduce diagonal matrices  ${\bf
G}=\diag(\gamma_1,..., \gamma_N)$ and ${\bf F}(\bq) =
\diag(|q_1|^2, |q_2|^2, \ldots, |q_N|^2)$, which describe the
dissipation and  the nonlinear part of the system, respectively.
Then the system (\ref{dynam_0}) can be  rewritten in the form
\begin{equation}
\label{dynam}
    i \dot{\bf q} = - [{\bf H} + {\bf F}(\bf q)]\bq, \quad \mbox{ ${\bf H} = {\bf K} +
i\bG$}.
\end{equation}
We search stationary nonlinear  modes  in the form $ {{\bf q}}(z)=
e^{ i {b} z}  {\bw}$, where  {$b$ is the propagation constant},
and $\bw = (w_1, w_2, \ldots, w_N)^T$   solves the stationary
DNLSE
\begin{equation}
\label{stat}
    b\bw = [{\bf H}+  {\bf F}(\bw)] \bw.
\end{equation}
Requiring the spectrum of  the
linear problem $ b\bw = {\bf H}\bw$  to be real,
 {which is necessary for all linear modes to be propagating, we
impose the constraint $\sum_{n=1}^N\gamma_n=0$}.

The matrix ${\bf H}$ is   $\PT$-symmetric  if it commutes with a
$\PT$ operator: $[\PT,\bH] =0$. Hereafter $\p$ is an orthogonal symmetric (and therefore Hermitian)
matrix, and $\T$ is element-wise complex conjugation: $\T \bq =
\bq^*$.
Using that  $\bH^\dag = \T \bH \T = \p
\bH \p$, we observe that the linear system $i \dot{\bf q} = -{\bf
H}\bq$
 admits an integral of motion (see also~\cite{Znojil})
$Q = \frac{1}{N} \langle \p \bq, \bq\rangle$,
 {where the inner product is defined as $\langle {\bf u}, {\bf v}\rangle = \sum_{n=1}^N v_n^*u_n $}.

\begin{figure}
\includegraphics[width=\columnwidth]{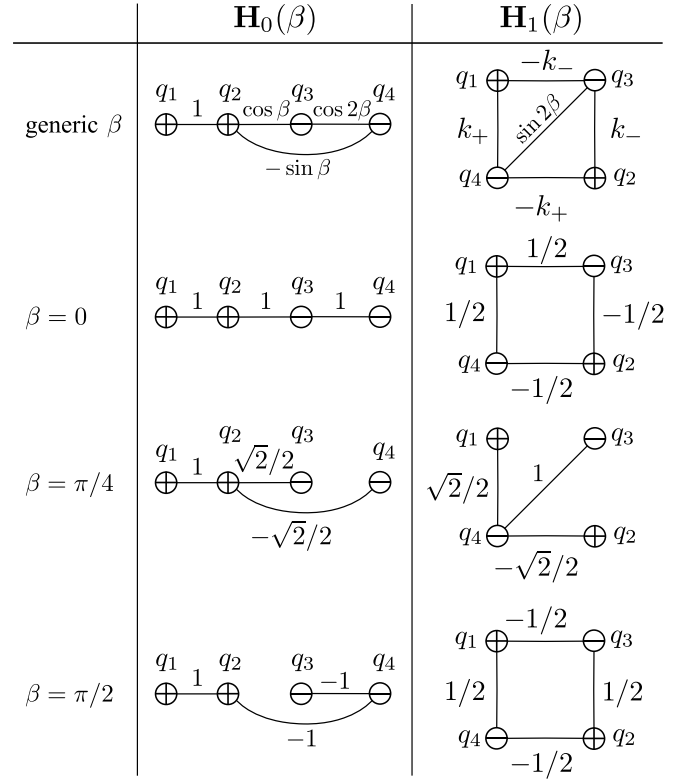}%
  \caption{Graph representation of the systems ${\bf H}_{0,1}(\beta)$ for generic and  particular values of $\beta$.  {In terms of the optical applications, the circles with ``$+$''  [or ``$-$'']
  represent
  waveguides with gain [or losses],  while the lines indicate the directions, along which the coupling of the field has to be arranged.} \label{fig-graphs}}
\end{figure}

Now we can specify the problem at hand: we consider the existence
and stability of nonlinear modes of  $\PT$-symmetric lattices
whose linear parts are related to each other by similarity
transformations, all having the nonlinearity of the on-site type.
Such a statement is natural for  arrays of optical waveguides,
since linear links among them can be arranged by assembling
waveguides in different geometries, while the dissipation/gain and
the nonlinearity are the characteristics of each particular
waveguide, which can be routinely controlled  {(see
also Fig.~\ref{fig-graphs})}. One of our main findings is that
{\em linearly equivalent   $\PT$-symmetric lattices result in
qualitatively different properties of their nonlinear extensions.}

\begin{figure}
\includegraphics[width=\columnwidth]{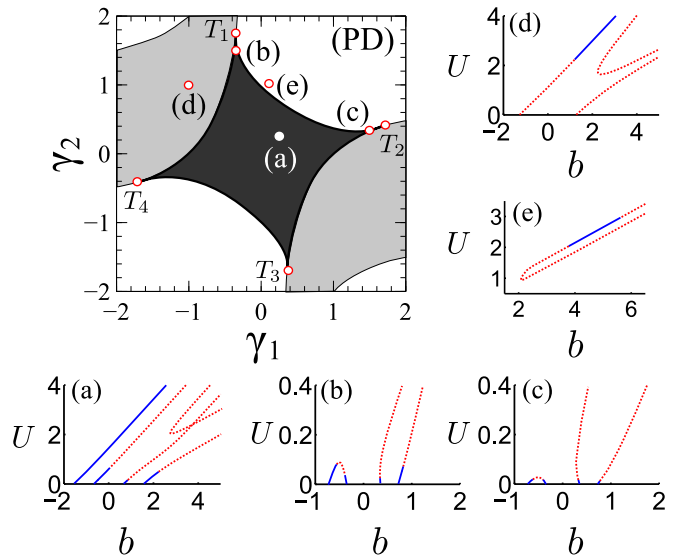}%
\caption{``Phase diagram'' (PD) for the linear
quadrimer $\bH_0$. The  dark-grey diamond-shaped domain
corresponds to unbroken $\PT$ symmetry; in the light-gray domains
there are two real and two complex eigenvalues. In the white
domains all eigenvalues are complex. In panels (a)--(e),
corresponding to the points (a)--(e) in the panel (PD), we show
families of nonlinear modes for:
  (a) $\gamma_{1,2}=0.25$;
  (b) $\gamma_{1} \approx -0.37$, $\gamma_2\approx 1.49$;
  (c) $\gamma_{1} \approx 1.49$, $\gamma_2\approx 0.36$;
  (d) $\gamma_{1,2}=\mp 1$;
  (e) $\gamma_{1} = 0.1$, $\gamma_2=0.95$.  Stable (unstable) modes are shown by solid blue (dashed red) lines.%
    }
 \label{fig-phasediag}
\end{figure}

\paragraph{$\PT$-symmetric ``quadrimer''--}
Since the gain and dissipation must compensate each other, the
simplest   models allowing for nontrivial distribution of the dissipation  have three or four waveguides. Below we
concentrate on a quadrimer, respectively setting $N=4$.  We start by revisiting
 recently considered in~\cite{Kevr2011}   system with the
next-neighbor interactions: $K_{nm} = \delta_{|n-m|, 1}$. The corresponding  matrix, which we denote as $\bH_0$, is
$\PT$-symmetric
with respect to  ${\p}_0 = \left( \begin {array}{cc}%
 {\bf 0}&\sigma_1\\%
 \sigma_1&{\bf 0}
\end {array} \right)$ (hereafter $\sigma_{1,2,3}$ are the Pauli matrices and ${\bf 0}$ is the $2\times 2$ zero matrix).
Depending on   particular values of $\gamma_{1,2}$, three
different situations are possible: (i) unbroken or \textit{exact}
$\PT$ symmetry, when  all the eigenvalues $\tb_n$, $n=1, ..., 4$,
of $\bH_0$ are real; (ii) broken $\PT$ symmetry with two real and
two complex conjugated eigenvalues (notice that this is possible
only if $\gamma_1 \neq \gamma_2$); (iii) broken $\PT$ symmetry
with all $\tb_n$ complex. Thus, the ``phase space'' $(\gamma_1,
\gamma_2)$ can be divided into  three domains as it is shown in
the phase diagram (PD) of Fig.~\ref{fig-phasediag}. A feature of
the phase diagram is the existence of the triple points $T_j$,
$j=1,\ldots, 4$,  where the three domains touch. The triple points
correspond to values $\gamma_{1,2}$ for which   $\tb_{n}=0$ for each $n=1,...,4$.
Depending on  how $\gamma_{1,2}$ change in vicinity of $T_j$,
either the $\PT$-symmetric phase  or one of the $\PT$ symmetry
broken phases arise.
\begin{figure}
\includegraphics[width=\columnwidth]{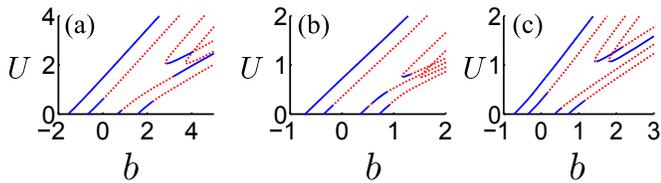}%
  \caption{Families of nonlinear modes of the Hermitian quadrimers,  {whose linear parts are described by} $\bH_H$,
  chosen   to have the same linear eigenvalues $\tb_n$ as the corresponding $\PT$-symmetric quadrimers  $\bH_0$ illustrated in the panel (a), (b), and  (c)  of Fig.~\ref{fig-phasediag}.
  \label{fig-Ubcons}}
\end{figure}

If   $\bH_0$ is    {exactly} $\PT$-symmetric, then its linear
eigenstates $\tbw$ are simultaneously  the eigenstates of the
corresponding $\PT$ operator, i.e. $\cP_0\T\tilde{\bw} =
\tilde{\bw}$ (up to irrelevant phase shift). It is natural to look
for nonlinear modes that possess the same property: $\cP_0\T\bw =
\bw$.  Therefore we require  $w_1 = w_4^*$, $w_2 = w_3^*$, which
reduces Eq.~(\ref{stat}) to
\bes%
\label{eq:reduced}
\begin{eqnarray}
b w_1 &=& w_2 + |w_1|^2w_1 + i\gamma_1 w_1, \label{eq:reda}\\
b w_2 &=& (w_1 + w_2^*) + |w_2|^2w_2 + i\gamma_2 w_2.
\label{eq:redb}
\end{eqnarray}
\ees
We represent $w_1 = W_1e^{i\Phi}$, where $W_1$ and $\Phi$ are
real. Then Eq.~(\ref{eq:reda}) gives $w_{2} = W_{2} e^{i\Phi}$, where
     $W_2 = {W_1}(b - W_1^2 -
    i\gamma_1)$ is complex, and from Eq.~(\ref{eq:redb}) we obtain
 $e^{-2i\phi} = f(W_1)$, where
  $
f(W_1) = %
 {
   \left(
   bW_2 - W_1 - |W_2|^2W_2 -
    i\gamma_2W_2
    \right)
    }/{W_2^*}
    $.
If a root of the equation $|f(W_1)|^2 = 1$ is found, then $w_1$
and $w_{2}$  can be readily obtained. Thus nonlinear   modes of
the quadrimer   correspond to the roots of a single equation
$|f(W_1)|^2 = 1$ with respect to one real unknown $W_1$. It is a
purely technical matter to
reduce  the latter  equation to: $P_8(W_1^2)=0$, where $P_8(\xi)$
is an eighth-degree polynomial with real coefficients. Each
positive root  of $P_8(\xi)$ corresponds to a nonlinear mode of
the quadrimer. Since the roots  depend continuously on $b$, the
nonlinear modes constitute \textit{continuous families} for fixed
parameters of the system~\cite{com3}.  {As it is
customary, such  families can be represented on the plane $(U, b)$
where $U=\frac{1}{4}\sum_{n=1}^4|w_n|^2$ is the total energy flow
in the array.} Panels (a)--(e) of Fig.~\ref{fig-phasediag}
illustrate
typical examples of the families, as well as linear stability of
the modes. When $\gamma_{1,2}$ belong to the  domain of unbroken
$\PT$ symmetry [see Fig.~\ref{fig-phasediag}~(a)], one observes
four families branching off from the  linear limit, i.e. from the
points $b=\tb_n$, $U=0$. In  Fig.~\ref{fig-phasediag}~(a) there
also exist families that can not be continued from the linear
limit. In  panels (b) and (c) we  also address the points that
belong to the domain of unbroken $\PT$ symmetry but are situated
closely to the triple points $T_{1,2}$. In these panels one
observes that after the bifurcation from the linear limit, all
four  families rapidly lose stability and  two of them cease to
exist if $U$ is sufficiently large. Comparing   panel (a) with
panels (b) and (c), we conclude that increase of $\gamma_{1,2}$,
i.e. approaching  the $\PT$ symmetry breaking boundary, is
unfavorable for existence and stability of the modes. However, the
most surprising fact, is that {\em stable  nonlinear modes can be
found in the domains of broken $\PT$ symmetry}. Both in panel~(d),
which addresses  the case when the  spectrum  consists of two real
and two complex eigenvalues, and in panel~(e),  i.e. when  all the
eigenvalues  are complex, one can find stable modes.

\paragraph{Hermitian quadrimer--}
If $\bH_0$ is exactly $\PT$-symmetric, then there exists a unitary
matrix ${\bf R}$, which transforms  $\bH_0$ to a Hermitian
matrix  $\bH_H$~\cite{Mostaf2003}: ${\bf R} {\bH_0}{\bf R}^{-1}
= \bH_H = \bH_H^\dag$.  {This means that in the linear limit the modes in  the  array with gain and losses described by $\bH_0$ have the same propagation constants as the modes in the array without gain and losses, which is described by $\bH_H$.}
 Hence, for any $\gamma_{1,2}$ lying in the domain of
unbroken $\PT$-symmetry of $\bH_0$, one can introduce  a new DNLSE
$i\dot{\bq}= -[\bH_H  +   {\bf F}(\bq)]\bq$ [c.f. (\ref{dynam})].
Following~\cite{Mostaf2003},  one can find $\bH_H$
explicitly and observe that  all its nonzero elements are real and
given by $\bH_{H, {12}}=\bH_{H, {21}}=h_1$, $\bH_{H, {14}}
=\bH_{H, {41}}=h_2$, $\bH_{H, {23}}=\bH_{H, {32}}=h_3$, with $h_j$
being dependent on $\gamma_{1,2}$. By construction, the matrix
$\bH_H$  has the same eigenvalues  as   $\bH_0$ for the  given
$\gamma_{1,2}$.

Unlike in the $\PT$-symmetric case, the modes  {of the nonlinear system with linear part described by $\bH_H$}
can be searched as  real-valued and   either even or odd,
i.e. solving the system
 $bw_1 = h_1w_2 \pm h_2w_1 +  w_1^3$, $bw_2 =
h_1w_1 \pm h_3w_2 + w_2^3$, where ``$+$'' (``$-$'')  stays for
even (odd) modes. This system  is
equivalent to a fourth-degree polynomial equation  with respect to
$w_1^2$. Families of even and odd nonlinear modes of the Hermitian
quadrimer are illustrated in Fig.~\ref{fig-Ubcons}. Comparing Fig.~\ref{fig-phasediag} and Fig.~\ref{fig-Ubcons}, we observe
that \textit{even if the  {matrices}  $\bH_0$ and $\bH_H$ have the
same eigenvalues,  {the respective nonlinear systems show considerable differences in the properties of modes}}. The most visible differences   are: (i) for the
Hermitian system, the families bifurcating from the linear limit
never close forming a saddle-node bifurcation [c.f. panels (b) and
(c) in Fig.~\ref{fig-phasediag} and Fig.~\ref{fig-Ubcons}]; (ii)
the leftmost  family of the Hermitian system is always stable;
(iii) in general, stable nonlinear  modes of  {Eq.~(\ref{dynam})} with $\bH_0$ and $\bH_H$ correspond to different values of the propagation constant $b$.

\paragraph{``Generalized'' quadrimer--}
Being of the dissipative nature, the considered above
$\PT$-symmetric quadrimer with linear part described by $\bH_0$  possesses a property, usually typical
for conservative systems -- for the given parameters of the system
(inter-site interactions $\bK$ and dissipation $\gamma_{1,2}$) its
nonlinear modes constitute continuous families rather than appear
as  isolated attractors. This peculiarity of nonlinear
$\PT$-symmetric systems was reported in several
studies~\cite{Christodoulides1,Kevr2011,PT-nonlin,PT-lin-nonlin}.
Here we argue  that existence of the continuous families of
nonlinear modes is not a typical property of $\PT$-symmetric
systems. Specifically, the nonlinear $\PT$-symmetric systems that
admit the  families of the modes appear as   ``isolated points''
in a continuous set of generic $\PT$-symmetric systems.

To this end we focus on the particular case $\gamma_1=\gamma_2 =
\gamma$, i.e.  ${\bf G} = \diag (\gamma, \gamma, -\gamma,
-\gamma)$, and introduce an one-parametric family of matrices
$\bH_0(\beta) = {\bf K}_0(\beta) + i{\bf G}$ with
\begin{equation*}
{\bf K}_0(\beta) = \left(\begin {array}{cccc}
0 & 1 & 0 & 0%
\\
1&0&\cos\beta&-\sin\beta\\
0&\cos\beta&\sin  2\beta  &\cos2\beta\\
0&-\sin\beta&\cos2\beta&-\sin2\beta   \end {array} \right),
\end{equation*}
and real parameter  $\beta$. One can ensure  that $\bH_0(\beta)$
is $\PT$-symmetric with respect to $
{\p}_0(\beta) = \left( \begin{array}{cc}%
 {\bf 0}&\rho(\beta)\\
\rho(\beta)&{\bf 0}%
 \end {array}\right)
$, where $\rho(\beta)=\cos\beta\ \sigma_1+\sin\beta\ \sigma_3$.
For $\beta=0$ the matrix $\bK_0(0)$   includes only the
next-neighbor interactions, i.e. $\bH_0(0)$ is merely the linear
part of the $\PT$-symmetric quadrimer studied above (with
$\gamma_1=\gamma_2$). Definition of $\bH_0(\beta)$ guarantees that
its  eigenvalues  do not depend on $\beta$. But the eigenvectors
of $\bH_0(\beta)$ do depend on $\beta$.

Next, using
$
{\bf M}  =  \left(\begin {array}{cc}
\mu & {\bf 0}\\%
{\bf 0}&\mu\end {array} \right)$,
where $\mu=\sigma_3+i\sigma_2$, one can generate a new matrix  ${\bH}_1(\beta) =  {\bf M}
\bH_0(\beta) {\bf M}^{-1}$, where ${\bH}_1(\beta) = {\bf
K}_1(\beta) + i\bG$,
\begin{equation*}
{\bf K}_1(\beta) =  \left(\begin {array}{cccc}
1 & 0 & -k_-&k_+\\%
%
0&-1&k_-&-k_+\\%
%
-k_-&k_-&\cos2\beta&\sin2\beta\\%
k_+&-k_+&\sin2\beta&-\cos2\beta  \end {array} \right),
\end{equation*}
and $k_{\pm} = \frac{\sqrt{2}}{2}\sin(\beta\pm \frac{\pi}{4})$.
Then $\bH_1(\beta)$ is $\PT$-symmetric with respect to $\p_1(\beta) = {\bf M}\p_0(\beta){\bf M}^{-1}$. Notice  that the    transformation  ${\bf M}$ does not affect the
dissipative component $i{\bf G}$, which is the same both for $\bH_0(\beta)$ and
$\bH_1(\beta)$. Obviously, the eigenvalues   of the matrix
$\bH_1(\beta)$ are the same  as for $\bH_0(\beta)$ and also do not depend on
$\beta$.

To give  better  physical insight into the  systems
$\bH_{0,1}(\beta)$,  in Fig.~\ref{fig-graphs}  we introduce their
weighted graph representation. The vertexes of graphs correspond
to the sites $q_n$ while the edges (lines) represent inter-site
coupling having weights equal to the values
of the respective matrix elements: e.g. a line between the vertexes
$q_1$ and $q_2$ corresponds to the elements $K_{1,2}= K_{2,1}$ of
the matrix ${\bf K}$. Each vertex is supplied by the
sign ``$+$'' or ``$-$'' corresponding to gain and dissipation. We
notice that the loop edges, which   describe the on-site
interactions $K_{n,n}$, are not shown  as being not relevant for the
present consideration.
The graph representation can be viewed also  as indication on how
one could place and connect the waveguides in an experiment in
order to obtain the desirable $\PT$-symmeric quadrimer. It is
worth noting that, say the bottom graph in left column [i.e.
$\bH_0(\pi/2)$] can be reshaped into the line distribution of the
waveguides similar to the    graph   $\bH_0(0)$.

\begin{figure}
\includegraphics[width=\columnwidth]{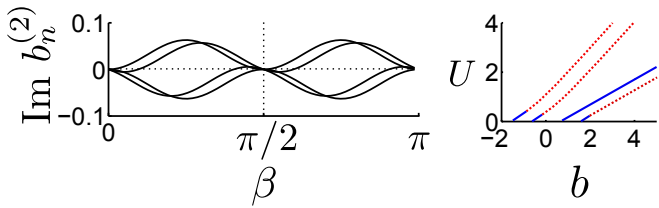}%
  \caption{$\IM \,b_n^{(2)}$ ($n=1,...,4$) {\it vs} $\beta$ for $\bH_0(\beta)$ (left panel) and the families of   modes of $\bH_1(0)$ for  $\gamma=0.25$  (right panel).
  \label{fig-b(beta)}}
\end{figure}

\paragraph{Existence of nonlinear modes--} Turning to
nonlinear properties of the  {arrays, whose linear links are described by}    $\bH_{0,1}(\beta)$, let us
suppose that  the $n$th eigenstate  of the underlying  linear problem $b\bw = \bH_{0,1}(\beta)\bw$
gives rise to a family of nonlinear modes. Then  in the vicinity
of the  bifurcation point the nonlinear modes can be described using
the expansion  $\bw = \varepsilon\tbw_n + o(\varepsilon)$, and  $b
= \tb_n + \varepsilon^2b_n^{(2)} + o(\varepsilon^2)$, where
$\varepsilon$ is a small parameter, $\tb_n$ and $\tbw_n$ are the eigenvalue and the
corresponding eigenvector of $\bH_0(\beta)$ [or $\bH_1(\beta)$].
The coefficient  $b_n^{(2)}$  can be readily found:
$
b_n^{(2)} =  {\langle {\bf F}(\tbw_n)\tbw_n,
\tbw_n^*\rangle}/{\langle \tbw_n, \tbw_n^*\rangle}$. This means that the  bifurcation of   nonlinear modes is possible
only if  $\IM\, b_n^{(2)}=0$ for all $n$ (we may conjecture that
this condition is also sufficient for existence of the modes, what
was observed in all our numerical simulations). The coefficient
$b_n^{(2)}$  is easily computable. In Fig.~\ref{fig-b(beta)} (left
panel)   $\IM\, b_n^{(2)}$ is plotted for $\bH_0(\beta)$.  Only at
$\beta=\beta_k=\pi k/2$ the coefficient $b_n^{(2)}$ becomes real
for all $n$ and the system $\bH_0(\beta)$ admits continuous
families of nonlinear modes, while for all other $\beta$ nonlinear
modes bifurcating from the linear limit were not found.

To understand peculiarity of the values $\beta_k$ we notice that relation $\PT\tbw_n =
\tbw_n$ ensures that the denominator
in the  formula for $b_n^{(2)}$
is real for any $\beta$: $\langle \tbw_n, \tbw_n^*\rangle = \langle \PT\tbw_n,
\T\tbw_n\rangle = \langle \T\tbw_n, \PT\tbw_n\rangle = \langle
\tbw_n^*, \tbw_n\rangle$.
Meanwhile  $\langle {\bf F}(\tbw_n)\tbw_n, \tbw_n^*\rangle$ can have nonzero
imaginary part. Then the reality of the coefficient
$b_n^{(2)}$ is ensured by an additional constraint
 $\PT({\bf F}(\tbw_n)\tbw_n) = {\bf F}(\tbw_n)\tbw_n$, which is satisfied only for $\beta=\beta_k$.

For the system $\bH_1(\beta)$ the situation is
similar --  the families of nonlinear modes
exist only for $\beta=\pi k/2$ where  $\IM\,b_n^{(2)}=0$ for all
$n$.  In   the right panel of Fig.~\ref{fig-b(beta)} we show families of nonlinear   modes  of the array whose linear part is described by   $\bH_1(0)$ with $\gamma=0.25$.  Comparing the  latter panel with  Fig.~\ref{fig-phasediag}~(a) (which also corresponds to $\gamma_{1,2}=0.25$), we again notice that,  whereas  the  corresponding arrays   have the same eigenvalues in the linear limit,  nonlinear modes of those  arrays  have   essentially different properties.

To conclude, we have considered nonlinear properties of different
$\PT$-symmetric lattices (discrete nonlinear Schr\"odinger
equations with gain and dissipation), whose linear parts are
related by similarity transformations preserving the spectrum.
{Such systems describe, in particular, arrays of optical
waveguides with either gain or losses, which are properly arranged
in the space.} Alternatively, a physical realization of the
described phenomenon is possible in arrays of Bose-Einstein
condensates loaded in multi-well potentials, provided the atoms
are eliminated from given wells and are condensed in the other
wells, simulating in this way losses and gain.

On the case example of a $\PT$-symmetric quadrimer  we have shown
that the spectral equivalence of the underlying linear systems
implies neither similarity of the nonlinear modes nor their
stability properties. We have found that the existence of
one-parametric families of nonlinear modes is not guaranteed by
the $\PT$ symmetry, and appears as a peculiarity of a system
rather than a general property. It was also found that the
stability of nonlinear modes is not directly related to the $\PT$
symmetry: stable nonlinear modes exist beyond the $\PT$ symmetry
breaking threshold. If the system includes  two different
dissipative coefficients, then the ``phase diagram'' of the
$\PT$-symmetric quadrimer allows for existence of ``triple''
points, where three different phases meet. Finally, we have shown
that use of graph representation of $\PT$-symmetric networks gives
{straightforward} indication on their possible experimental design
in optics, and provides graphical illustration of linearly
equivalent networks.

\smallskip

Authors acknowledge support of the FCT (Portugal) grants:    SFRH/BPD/64835/2009, PTDC/FIS/112624/2009, and PEst-OE/FIS/UI0618/2011.


\begin{thebibliography}{99}

\bibitem{diss_solit} see e.g. \textit{Dissipative Solitons},
eds. N.~Akhmediev and A.~Ankiewicz (Springer-Verlag, 2005);  Focus Issue: \emph{Disspative Localized Structures in Exteded
Systems}, Chaos \textbf{17}, (2007).

\bibitem{waveguide}  Y. Chen, A. W. Snyder, and D. N. Pain,
IEEE J. Quant. Electron. {\bf 28}, 239 (1992).

\bibitem{disc_opt} F. Lederer, G. I. Stegeman, D. N. Christodoulides, G. Assanto, M. Segev, and Y. Silberberg, Phys. Rep. {\bf 463}, 1
(2008).

\bibitem{biophysics} A. Scott, \textit{Nonlinear Science. Emergence and Dynamics of Coherent Structures} (Oxford, University Press,
1999).

\bibitem{BEC} P. G. Kevrekidis and  Frantzeskakis, Mod. Phys. Lett. B {\bf 18}, 173 (2004); V. A. Brazhnyi, V. V. Konotop, Mod. Phys. Lett. B {\bf 18}, 627 (2004); O. Morsch and M. Oberthaler, Rev. Mod. Phys. {\bf 78}, 179 (2006).



\bibitem{general} P. G. Kevrekidis,  \textit{The Discrete Nonlinear Schrödinger Equation}  (Springer, Berlin Heidelberg
2009).

\bibitem{Bender} C. M. Bender  and S. Boettcher,  Phys. Rev. Lett. {\bf 80}, 5243 (1998).

\bibitem{quantum} C. M. Bender, S. Boettcher, and P. N. Meisinger, J. Math. Phys. {\bf 40}, 2201 (1999).

\bibitem{Muga} A. Ruschhaupt, F. Delgado, J. G. Muga, J. Phys. A: Math. Gen. {\bf 38}, L171  (2005).

\bibitem{Markis} K. G. Makris, R. El-Ganainy,  D. N. Christodoulides, and Z. H. Musslimani, Phys. Rev. Lett. {\bf 100}, 103904 (2008); S. Klaiman, U. G\"unther, and N. Moiseyev, Phys. Rev. Lett. {\bf 101} 080402 (2008).

\bibitem{experiment}
C. E. R\"uter, K. G.  Makris, R. El-Ganainy,
D. N. Christodoulides,  M. Segev, and   D. Kip, Nature
Phys. {\bf 6} {192} {(2010)}.


\bibitem{Bender_nonlin} C. M. Bender, D. C. Brody, and H. F. Jones, Phys. Rev. D {\bf 70}, 025001 (2004).

\bibitem{Christodoulides1} Z. H. Musslimani, K. G. Makris, R. El-Ganainy, and D. N. Christodoulides
Phys. Rev. Lett. {\bf 100}, 030402, (2008).

\bibitem{defect}
Xing Zhu, Hong Wang, Li-Xian Zheng, Huagang Li, and Ying-Ji He, Opt. Lett. \textbf{36}, 2680 (2011).

\bibitem{coupler_stat} H. Ramezani, T. Kottos, R. El-Ganainy, and D. N.
Christodoulides, Phys. Rev. A {\bf 82}, 043803 (2010); A. A. Sukhorukov, Z. Xu, and Yu. S. Kivshar, Phys. Rev. A {\bf 82},
043818 (2010).


\bibitem{Kevr2011}
K. Li and P. G. Kevrekidis,  Phys. Rev. E  \textbf{83}, 066608
(2011).

\bibitem{coupler_solit} R. Driben and B. A. Malomed Opt. Lett. {\bf 36}, 4323 (2011); F. Kh. Abdullaev, V. V. Konotop, M. \"{O}gren, and
M. P. S\o rensen Opt. Lett.  {\bf 36}, 4566 (2011).


\bibitem{PT-nonlin} F. Kh. Abdullaev, Y. V. Kartashov, V. V. Konotop, and D. A. Zezyulin, Phys. Rev. A  {\bf 83}  041805(R), {(2011)}; D. A. Zezyulin, Y. V. Kartashov, V. V. Konotop, Europhys. Lett. {\bf 96}, 64003 (2011).

\bibitem{PT-lin-nonlin}  A. E. Miroshnichenko,  B. A. Malomed,  and  Yu. S.
 Kivshar, Phys. Rev. A  {\bf 84}, {012123}  (2011);  Y. He, X. Zhu, D. Mihalache, J. Liu, and Z. Chen,    
 Phys. Rev. A  {\bf 85}, 013831 (2012).


\bibitem{Darboux} F. Cannata, G. Junker, and J. Trost, Phys. Lett. A {\bf 246}, 219
(1998).

\bibitem{Mostaf2003} A. Mostafazadeh,  J.  Math. Phys.   \textbf{43}, 205 (2002);
J. Phys. A: Math. Gen.  \textbf{36}, 7081 (2003).

\bibitem{Znojil} B. Bagchi, C. Quesne, and M. Znojil, Mod. Phys. Lett. A \textbf{16}, 2047 (2001).

\bibitem{com3}{The continuous family of solutions presented herein is an additional one to the solutions
found in \cite{Kevr2011} for $\gamma_1=\gamma_2$, which required
that the parameters of the system are  inter-related. These two
types of solutions are complementary in the complete set of
possible standing wave solutions in this special case of equal
values of the quadrimer gain/loss parameters. }

\end{thebibliography}
\end{document}